# Analog input layer for optical reservoir computers

**François Duport[1], Akram Akrout[2], Anteo Smerieri[1], Marc Haelterman[1], and Serge Massar[2]**
[1]OPERA-Photonique, CP194/5, Université libre de Bruxelles (ULB),
Av. F. D. Roosevelt 50, B-1050 Bruxelles
[2]Laboratoire d'Information Quantique, CP225, Université libre de Bruxelles (ULB),
Av. F. D. Roosevelt 50, B-1050 Bruxelles

E-mail: francois.duport@ulb.ac.be

**Abstract.** Reservoir computing is an information processing technique, derived from the theory of neural networks, which is easy to implement in hardware. Several reservoir computer hardware implementations have been realized recently with performance comparable to digital implementations, which demonstrated the potential of reservoir computing for ultrahigh bandwidth signal processing tasks. In all these implementations however the signal pre-processing necessary to efficiently address the reservoir was performed digitally. Here we show how this digital pre-processing can be replaced by an analog input layer. We study both numerically and experimentally to what extent the pre-processing can be replaced by a modulation of the input signal by either a single sine-wave or by a sum of two sine functions, since harmonic oscillations are particularly easy to generate in hardware. We find that the modulation by a single sine gives performance worse than state of the art. On the other hand, on many -but not all- tasks, the modulation by two sines gives performances comparable to the state of the art. The present work thus represents an important step towards fully autonomous, ultrahigh bandwidth analog reservoir computers.

## 1.Introduction

Reservoir computing is a recently introduced, highly efficient bio-inspired approach for processing time dependent data [1,2] (see [3] for a review on the subject). A reservoir computer is essentially a nonlinear recurrent dynamical system coupled to a linear "input" layer and a linear "output" layer, used to perform complex computation on inputs encoded as time series, see figure 1 for a schematic depiction. Several hardware implementations of a reservoir computer have been realised recently with performance comparable to digital implementations. These include electronic [4], optoelectronic [5,6] and optical [7,8] implementations. These analog reservoir computers are all based on a time multiplexing architecture, consisting of a single nonlinear node and a delay line, introduced in [4,9].

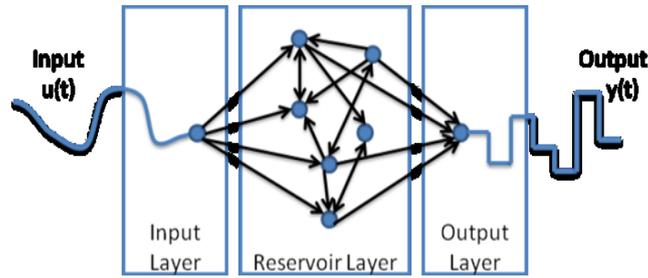

**Figure 1.** Schematic of a reservoir computer, including the input, reservoir, and output layers.

In all these works only the reservoir itself has been physically implemented, while the functions of the linear input and output layers were realized by digital pre- and post-processing, which makes those implementations inherently slow. In order to have a complete high-bandwidth hardware implementation of the reservoir computing concept it is necessary to include analog implementations of both the input and output layers. Moreover, it is worth noting that besides the development of fully autonomous reservoirs, this improvement will allow multiple reservoirs to be coupled together, or allow new regimes of operation such as pattern generation based on teacher forcing [2] or FORCE training [10].

The importance of the input layer is underlined by two recent works that study how, by modifying the input layer, one can improve performance of reservoir computers based on the single nonlinear node and delay line architecture. In [11] it was indeed shown how multilevel input masks can increase the resistance to noise, while a deterministic procedure to choose an input mask with a minimum number of nodes was proposed in [12].

Recently we reported the first analog implementation of the output layer [13]. Here, we report an analog implementation of the input layer for reservoirs, adapted to our opto-electronic [5] and all-optical [7] experiments. This is realised by modulating the optical input signal by a periodic function using standard telecommunication components. The present work thus demonstrates that it is not necessary to carry out any digital preprocessing of the input signal and that reservoirs can directly process input signals provided by an external source.

When using the single nonlinear node and delay line architecture, the analog input layer consists in modulating the input signal using a periodic function. In typical implementations of reservoir computing, this periodic function, called the input mask, is a random function. However random functions can be difficult and/or costly to implement. Indeed specifying a random function requires a large number of parameters, and therefore its hardware implementation requires expensive and high bandwidth electronics, such as a high speed Arbitrary Waveform Generator. Previous studies on the internal connectivity of reservoir computers have shown that one can use, with no performance degradation [14], a simple deterministic structure for the reservoir layer, instead of the randomly generated layers that are traditionally used. This holds provided that the internal structure of the reservoir is complex enough, while overly simple structures tend to give worse performances [15]. The second aim of this paper is to find out if we can apply the same concept to the input mask, and find a deterministic way of building an input mask that is both easy to implement in hardware and complex enough to give good computational performance.

We have therefore tried as input mask a simple function, namely a sine function. Indeed being the output of an oscillator, harmonic functions are the simplest functions to generate in hardware. Our simulations showed that the sine function gives reasonable performance on a standardized benchmark task (equalisation of a nonlinear communication channel), but not as good as random functions. In view of this, we made the input mask slightly more complex, and took it to be the sum of two sine functions with different periods. After optimising over the relevant discrete parameters (number of nodes, periods of the two sines, etc.), the performance on this benchmark task became comparable to random functions. The analog input layer consisting of the sum of two sines was then implemented in experiment.

We tested the analog input layer on both our opto-electronic [5] and all-optical [7] reservoirs. We obtained in both cases excellent performances on the nonlinear channel equalisation task. We also obtained good results in other tasks for which the input mask was not specifically optimized, like the computation of a nonlinear autoregressive moving averages or some instances of time series prediction based on real data. However for some other tasks, such as the prediction of a chaotic time series, the simple analog mask was not as effective as the traditional random mask. The implications of these results are discussed in the conclusion. In future work we intend to combine the analog input layer studied here and the analog output layer proposed in [13] to demonstrate the first fully autonomous reservoir computer.

## 2. Theoretical foundations

*2.1. Reservoir Computing Basics*

The core of a reservoir computer is a nonlinear, complex dynamical system, called "reservoir layer". It can be seen as a collection of internal variables, often called "nodes". The evolution, in discrete time $n$, of the states $x_i(n)$ of the internal variables when a signal is sent to the reservoir is given by:

$$x_i(n) = f_{NL}\left(\alpha \sum_{j=1}^{N} W_{ij}^{res} x_j(n-1) + \beta m_i u(n)\right), \qquad i = 1, 2, \ldots, N \qquad (1)$$

where $N$ is the number of variables, $f_{NL}$ is a nonlinear function, $u(n)$ is the reservoir input signal at time n, and $W_{ij}^{res}$ are the connection coefficients that describe the network topology. The feedback gain $\alpha$ and the input gain $\beta$ are parameters that regulate the dynamics of the reservoir.

The node-dependent coefficients $m_i$ are called collectively the input mask: they can be seen as a linear input layer that distributes different versions of the same input $u(n)$ to the internal nodes $x_i$. The purpose of the input mask is to break any symmetry between the temporal evolution of internal variables $x_i(n)$, enriching the dynamics and the computational capabilities of the reservoir.

The purpose of a Reservoir Computer is to carry out computation on the input signal $u(n)$. The result $y(n)$ of the computation is built in the output layer. It is given by a linear combination of the node states:

$$y(n) = \sum_i W_i^{out} x_i(n) \qquad (2)$$

The readout weights $W_i^{out}$ are chosen to minimize the squared difference between the obtained output $y(n)$ and the desired one $d(n)$. As building the output from the reservoir states is a linear operation, choosing the readout weights $W_i^{out}$ is a straightforward and computationally inexpensive operation; this ease of training is one of the main advantages of Reservoir Computing.

*2.2. Time multiplexing*

In most digital implementations the connection matrix $W_{ij}^{res}$ in equation (1) is randomly generated. However, it has been shown that even a simple interconnection matrix with nearest neighbour coupling can lead to excellent results [14]. For our experimental realization of hardware based reservoir computing we use an architecture where each node state $x_i$ is only connected to a neighbouring node state $x_{i-k}$, where $k$ is the offset. Such a configuration is described in detail in [4,

5]. It presents significant practical advantages, as it can be realised with just a single nonlinear element and a delay line, using time multiplexing of the reservoir states. Note that in our previous works [5, 7] the offset parameter $1 \leq k \leq N$ was always taken equal to $k = 1$. As discussed below, we take it here to be a free parameter that can be optimised.

To enforce this configuration experimentally, the output $x(t)$ of the nonlinear unit is fed back into its input after a delay $T$, and added to the product $m(t)u(t)$ of the continuous input signal $u(t)$ and the continuous mask $m(t)$. This yields the following dynamical system:

$$x(t) = f_{NL}\big(\alpha x(t - T) + \beta m(t)u(t)\big) \tag{3}$$

Here $f_{NL}$ is the transfer function of the nonlinear unit. The continuous time input $u(t)$, represented by the blue dashed line in figure 2, is obtained from the discrete time input *u(n)* by a sample and hold procedure:

$$u(t) = u(n) \quad \text{for} \quad nT' \leq t < (n+1)T' \tag{4}$$

where the hold time $T'$ of the input may be taken different from the delay $T$ of the dynamical system. The continuous time input mask $m(t)$ is a periodic function of period $T'$, $m(t + T') = m(t)$; this translates, for a time multiplexed reservoir computer, the requirement that the input mask coefficients $m_i$ should depend only on the node addressed and not on the time (see Eq. 1).

From the continuous signal $x(t)$ one can then define $N$ discrete states $x_i(n)$ by defining a "state duration" $\theta = T'/N$ and using the discretisation:

$$x_i(n) = x(t) \quad \text{for} \quad nT' + (i-1)\theta \leq t < nT' + i\theta \tag{5}$$

and defining a discrete time input mask $m_i$ by a sample and hold procedure (the red dots in figure 2):

$$m(t) = m_i \quad \text{for} \quad i\theta \leq t < (i+1)\theta \tag{6}$$

In order to obtain a rich enough dynamical behaviour for good performance as a reservoir computer, two procedures have been used in the literature. In [4,6] the synchronized regime $T = T'$ was chosen, and the transfer function $f_{NL}$ contained a low pass filter that coupled successive nodes together, whereas in our works [5,7] we adopted the unsynchronized regime $T > T'$ with

$$T - T' = k\theta = kT'/N, \quad k \in \{1, \ldots, N - 1\} \tag{7}$$

and an essentially instantaneous transfer function $f_{NL}$. It can then be shown [5] that the corresponding discrete model is

$$\begin{aligned} x_i(n) &= f_{NL}\big(\alpha x_{i-k}(n-1) + \beta m_i u(n)\big), \quad i = k+1, \ldots, N \\ x_i(n) &= f_{NL}\big(\alpha x_{i-k+N}(n-2) + \beta m_i u(n)\big), \quad i = 1, \ldots, k \end{aligned} \tag{8}$$

In the present work we use the latter approach, with $T - T' = k\theta$ and $f_{NL}$ an instantaneous transfer function.

*2.3. Analog Input Mask design*

In all earlier works the input mask $m_i$ was chosen at random, for instance from the discrete set {-1,+1}, or from the uniform distribution over [-1,+1]. The importance of the input mask is that it breaks the symmetry of the system and thereby enriches its dynamics. Indeed, thanks to the mask each node $x_i(t)$ has a different dependence on the previous inputs, thereby providing a rich set of functions from which to build the output. The fact that a randomly chosen mask gives good performance indicates that most input masks give good performance. Indeed it implies that only a set of small measure gives bad performance, and also that only a set of small measure will give much better performance than the average, see [12] for a discussion of this point. The only known requirement of the input mask is that its value should depend only on the node addressed, which is achieved in time multiplexing configurations by imposing a periodicity of period T'. On the other hand random functions with the right periodicity are costly to implement in hardware. Typically they are specified by many parameters (they have high entropy), implying that the hardware generator should have high complexity. Furthermore in all previous experiments [5-9] the input mask m(t) was a step function, as in equation (6), which requires a high bandwidth, and therefore an expensive arbitrary waveform generator. An example of a step function input mask is the one represented by the green line in the upper panel of figure 2.

Here we propose to use as input mask $m(t)$ as simple a function as possible, in analogy with previous works focused on simple deterministic structures of the nonlinear reservoir layer [14]. To this end we have concentrated on sine functions, as these can be easily generated in hardware using an oscillator. Initially we considered a continuous mask $m(t)$ that consists of a single sine of frequency $F_1/T'$:

$$m(t) = \sin\left(\frac{2\pi t}{T'} F_1\right), \text{ with } F_1 \text{ an integer.} \tag{9}$$

Its discrete analogue is

$$m_i = \sin\left(\frac{2\pi i}{N} F_1\right), \text{ with } i = 1, 2, \ldots, N \text{ and } F_1 \in \{1, \ldots, N\} \tag{10}$$

However, as the results of simulations discussed in Section 4 of this paper show, we did not obtain good performances using the input mask described by equation (9). One motivation might be that due to its simplicity the input mask fails to provide dynamics of the reservoir layer that are rich enough to leverage the high dimensionality of the reservoir layer. The theory of reservoir computing shows that to obtain a good reservoir it is necessary (but may not be sufficient) for the internal states of the reservoir to be linearly independent deterministic functions of the previous inputs [16]. It is obvious that if the mask is too simple (e.g. a constant mask), then the internal variables will not be linearly independent.

We therefore considered a slightly more complicated continuous mask $m(t)$ that consists of the sum of two sine functions of frequencies $F_1/T'$ and $F_2/T'$ respectively:

$$m(t) = \sin\left(\frac{2\pi t}{T'} F_1\right) + \sin\left(\frac{2\pi t}{T'} F_2\right), \text{ with } F_1, F_2 \text{ integers.} \tag{11}$$

Its discrete analogue is

$$m_i = \sin\left(\frac{2\pi i}{N} F_1\right) + \sin\left(\frac{2\pi i}{N} F_2\right), \quad \text{with: } i = 1, 2, \ldots, N \text{ and } F_1, F_2 \in \{1, \ldots, N\} \tag{12}$$

The product $m(t)u(t)$ of the input signal and of a sinusoidal mask is shown in figure 2 as the green line, and compared to the same product when a randomly generated, stepwise constant input mask is used on the same input. The raw input $u(t)$ to the reservoir computer is shown as a blue dashed line, while the discretized values $m_i u(t)$ representing the input to each node are represented as red dots.

By taking $N = 53$ and optimizing over the discrete parameters $F_1$, $F_2$ and $k = (T - T')/\theta$, as well as the continuous parameters $\alpha$ and $\beta$, we obtained performance similar to those obtained with a random mask, and therefore did not proceed to further increase the input mask complexity.

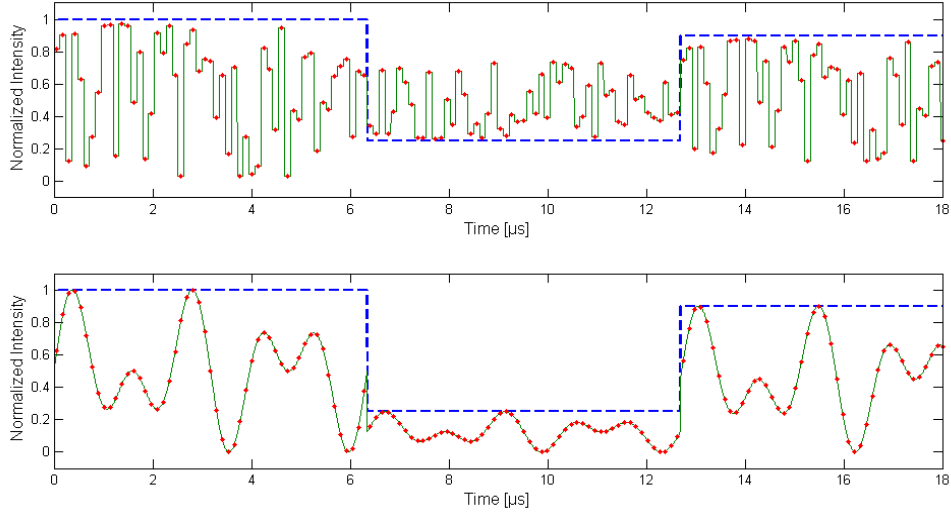

**Figure 2.** The input $u(t)$, dashed in blue, and the product of input and mask $m(t)u(t)$, in green, that is injected into the reservoir layer. Upper panel: standard random mask. Lower panel: input mask consisting of the sum of two sines, with $F_1 = 3$ and $F_2 = 5$. The input $u(t)$ is a step function constant over intervals of duration $T' = 6.2\ \mu s$. The random mask is drawn from the uniform distribution over [-1, +1]; it is a step function constant over intervals of duration $\theta = T'/N$ and periodic of period $T'$. The number of nodes is N=53. The red dots represent the values $m_i u(n)$ used in the discrete model.

## 3. Hardware implementation

### 3.1. Input layer

We have added an analog input layer to both our opto-electronic [5] and all-optical [7] reservoirs (see these references for additional information, and Figs. 3 and 4 for schematics of the experimental setups). Both these systems are fibre based and operate in the telecom C-band. In order to avoid any interference between the optical signal in the reservoirs and the analog optical input signal, we use an incoherent light source.

The input layer for both experiments is built as follows. The optical source is a continuous wave Super Luminescent Light Emitting Diode (SLED) (output centred on 1560nm, with a FWHM of 40nm). The input $u(t)$ is encoded by modulating light intensity using a voltage driven intensity modulator (Lithium Niobate Mach Zender interferometer) in such a way that the light intensity entering the system is proportional to $u(t)$). The input mask is realized by modulating the light intensity using a second intensity modulator to produce a light intensity proportional to $m(t)u(t)$.

Because they are available and already integrated with the rest of the experiment, we used two 200 MSamples/s Arbitrary Waveform Generators (AWG) to drive the two MZ (although slower and less versatile equipment could of course also have been used). The 200 MSamples/s rate of the AWGs is much faster than all other time scales in our systems, so that $m(t)$ can be considered a continuous function of time.

Finally the input intensity is adjusted using a variable attenuator, thereby providing access to the input gain $\beta$.

Note that the encoding of the input and of the input mask are two separate operations, contrary to all previous realisations where the product $m(t)u(t)$ was first realized using digital pre-processing, and then encoded in a single step. Conceptually, this corresponds to the situation that one would encounter in practice in which the input signal is provided by an external source. While we used an AWG to drive the second Mach-Zehnder, responsible for the input mask implementation, we did so only to be able to optimize the values of $F_1$ and $F_2$. Should this architecture be used in a real world application, no waveform generator would be necessary at all: one can find the optimal values of $F_1$ and $F_2$ beforehand, even in simulation only, and only then build a simple oscillator circuit that provides the right frequencies, while of course the input signal $u(t)$ would be provided by an external source.

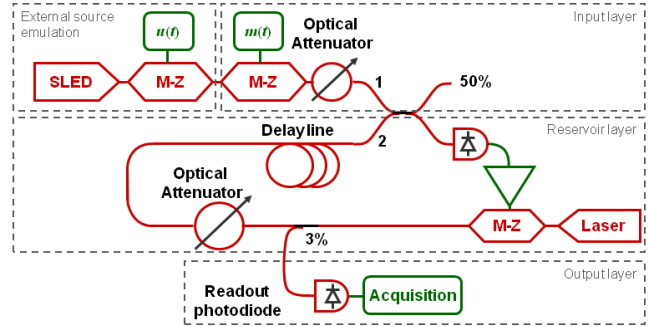

**Figure 3.** Optoelectronic implementation. The external source is emulated by a MZ modulating the output of a SLED to give the input signal $u(t)$. The input layer then provides the multiplication by the mask $m(t)$ and an input gain control via a tuneable attenuator. The reservoir layer consists of a laser, a MZ modulator, a variable attenuator, a fibre spool, a coupler to inject the input, a photodiode and an amplifier. The output of the amplifier is used to drive the MZ modulator. The output layer consists of a photodiode and digitizer. Optical components are depicted in red, electronic components in green.

*3.2. Reservoir layer*
The reservoir consists of a delay loop and a nonlinearity.
*Opto-electronic implementation*: In this case, the nonlinearity consists of a voltage driven MZ intensity modulator fed by a continuous light source. This implements a sine nonlinearity. The delay loop consists of a fibre spool and a tuneable optical attenuator. Its duration is of 8.5 $\mu s$, which leads to a value of $\theta = 157\ ns$. The latter is used to adjust the strength of the feedback gain $\alpha$ by tuning the light intensity inside the fibre spool. Tuning the feedback gain is necessary to optimize the performance of the reservoir.
The light intensity at the output of the fibre spool is combined with the light coming from the input layer and converted into a voltage using a photodiode.
*All-optical implementation*: In this case the nonlinearity is realized by the saturation nonlinearity of a Semiconductor Optical Amplifier (SOA). In addition an isolator is inserted in the fibre loop, and bandpass filters are inserted in the input and reservoir layers to remove spontaneous emission noise. In this case the delay loop has duration $T = 7.9\mu$ $(\theta = 146\ ns)$

*3.3. Output Layer*
Part of the signal propagating in the delay loop is split off and sent to the readout photodiode. The light intensity is thus converted to a voltage and recorded by a digitizer operating at 200MSamples/s. Variables $x_i(n)$, $i = 1, ..., N$ are obtained by averaging the recorded value over the duration $\theta$. The output $y(n)$ is taken to be a linear combination of these $N$ variables with weights that are optimized, see equation (2).

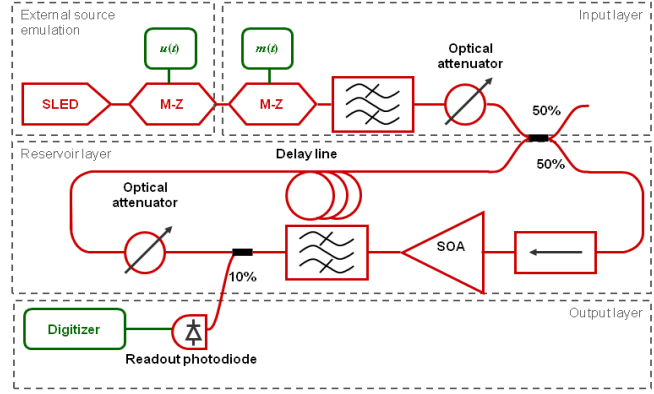

**Figure 4.** All-optical implementation. The input and output layers are identical to those of figure 3, except for the addition of a bandpass filter to reduce the spectrum of the SLED. The reservoir layer consists of an isolator, an SOA (that realises the nonlinearity), a bandpass filter, a variable optical attenuator and a fiber spool. Optical components are depicted in red, electronic components in green.

## 4. Results

*4.1. Tasks*
We have tested our reservoirs with several different tasks that were already shown to be of interest for the reservoir computing community, namely nonlinear channel equalization, computation of nonlinear autoregressive moving averages, prediction of chaotic time series and real world measurement prediction. Specifically, we used the nonlinear channel equalization task to compare between different possible choices of input mask shapes and discrete parameters in the network; once such choices were made, the other tasks were used as validation of the resulting architecture. We also tried to characterize the reservoirs by measuring their capabilities of recalling linear and nonlinear functions of past inputs. What follows is a description of the computational tasks used.

*4.1.1 Channel equalization*
The nonlinear channel equalization involves a simplified model of wireless transmission in which the input signal travels through multiple paths and is collected by a noisy and nonlinear receiver. This problem is often used as a benchmark task in the Reservoir Computing community, see e.g. [2,5,7,14]. The signal transmitted through the channel consists of a sequence of symbols $d(n)$ taken from the set {-3, -1, 1, 3}. Firstly, from the original sequence, we create a sequence $q(n)$ that takes into account multiple reflections in the wireless channel

$$\begin{aligned} q(n) = \; & 0.08d(n+2) - 0.12d(n+1) + d(n) + 0.18d(n-1) \\ & -0.1d(n-2) + 0.091d(n-3) - 0.05d(n-4) \\ & +0.04d(n-5) + 0.03d(n-6) + 0.01d(n-7) \end{aligned} \quad (12)$$

From the resulting series $q(n)$ we create a sequence $u(n)$ that takes into account the distortion in the receiver and the presence of noise:

$$u(n) = q(n) + 0.036q^2(n) + 0.011q^3(n) + noise \quad (13)$$

where the noise term can be tuned to achieve a Signal to Noise Ratio (SNR) between 12 dB and 32 dB. The task for the reservoir is then, given the sequence $u(n)$, to find the target sequence $d(n)$. The quality of the reservoir performance is measured by the Symbol Error Rate (SER), the proportion of input symbols $d(n)$ that are incorrectly reconstructed.

We used this task as an initial benchmark, when looking for an input mask that could give good results, and used the rest of the tasks described here to validate the choice of the input mask.

*4.1.2 Nonlinear AutoRegressive Moving Average (NARMA10)*

The nonlinear autoregressive moving average task was first introduced by [17] and then frequently employed as a benchmark for reservoir computers. It consists, given an input sequence $u(n)$ of samples randomly drawn from the interval [0, 0.5], in calculating the desired output $d(n)$, defined as

$$d(n) = 0.3\, d(n-1) + 0.05 d(n-1)\left(\sum_{i=1}^{10} d(n-i)\right) + \\ +1.5 u(n-10)u(n-1) + 0.1 \tag{14}$$

The performance is usually measured by the Normalized Mean Square Error (NMSE) which, if $d(n)$ is the target output and $y(n)$ is the actual output of the reservoir, is given by

$$NMSE = \frac{1}{N}\sum_{n=1}^{N} \frac{\big(d(n)-y(n)\big)^2}{var(d)} \tag{15}$$

*4.1.3 Memory capacities*

This basic task is generally used to gather some insights on the one fundamental property of the reservoir, namely its ability to recall previous inputs or some basic function of their value [16,18].

For this task, the input $u(n)$ to the reservoir is a sequence of symbols randomly drawn from an uniform distribution over the interval [-1,1]. It is then possible to use Legendre polynomials to evaluate the processing capabilities of the reservoirs over an orthogonal set of nonlinear functions as it was defined in [16].

To measure the i-th *linear memory capacity* $C_i$, one sets the target output as $d(n) = u(n-i)$, records the reservoir output $y_i(n)$, calculates the NMSE value $N_i$ and then defines $C_i = 1 - N_i$. The value $C_i$ is then a measure of how well the reservoir can hold in its memory the value of an input presented $i$ timesteps in the past. One can then define a total linear memory capacity $C = \sum_i C_i$, which is broadly speaking the amount of linear memory available for the system.

By analogy with the linear memory capacities, one can then define the nonlinear *cross memory capacity* on the same input $u(n)$ by changing the target function to be $d_{ij}(n) = u(n-i)u(n-j)$, for $j > i$, which measures the capability of the reservoir to recall a nonlinear function of past inputs. The special case $i = j$, called the *quadratic memory capacity*, is better covered by the target function $d_i(n) = 3u^2(n-i) - 1$, which is the Legendre polynomial of the second order over the interval [-1, 1] and insures that the quantity measured does not overlap with the linear memory capacity discussed above. The total memory capacity, i.e. the sum over all the previous times $i$ and $j$ of the linear, quadratic and cross memory capacity, is always bounded by the total number $N$ of nodes in the reservoir layer [16].

*4.1.4 Radar signal prediction*

In this case, the signal $u(n)$ is a radar signal, backscattered from the ocean surface and collected by the McMaster University IPX radar [19]. The task for the reservoir, when the input $u(n)$ is being fed to the reservoir, is then to predict the value of $d(n) = u(n + p)$, where $p$ is called the prediction horizon. The performance is measured usually by the NMSE.

It should be noted that the available datasets provide the complex envelope of the signal received by the radar. As the architecture of our input mask only allows for inputs that are monodimensional, real valued time series, we feed the reservoir with either the real or the imaginary part of the data, and we then average over the two experiments the resulting NMSE value. There were no significant differences in performance when using the real or the imaginary part of the data.

*4.1.5 Mackey-Glass time series prediction*
The Mackey-Glass time series is the solution to the nonlinear time delayed differential equation

$$\frac{du}{dt} = a \frac{u(t - \tau)}{1 + u(t - \tau)^c} - bu \qquad (16)$$

where $a, b, \tau$ and $n$ are parameters that regulate how chaotic the resulting solution is. We chose $a = 2$, $b = 1$, $\tau = 17$, and $c = 10$, in order to achieve mildly chaotic behaviour, and then integrated the resulting equation using a Runge-Kutta method, starting from a random value of $x$ in the interval $(0,1)$. The resulting time series, after a washout period, was sampled with unit sampling period to create the input to the reservoir $u(n)$. Here, as for the radar time series, the goal is to predict the signal $p$ steps into the future, so we define the desired output of the reservoir $d(n)$ as $d(n) = u(n + p)$. Like in the radar task, the performance is measured using the NMSE.

*4.2. Numerical simulations*
Before carrying out experiments, we checked the performance of the opto-electronic setup with an analog input mask by running numerical simulations on the Channel equalization task. We ran two different simulation sets. The first is the simplified discrete model given by equation (8). It neglects all bandpass and noise effects of the experimental setup. In the second simulation, the behaviour of each component is modelled realistically in continuous time, including noise and filtering behaviour. For brevity we will refer to them as the "discrete" and the "continuous" simulations, respectively.
Both simulations are described in detail in the supplementary materials of [5]. They allow for an efficient optimization of the reservoir parameters and validation of the experimental results. In general for each task, the different parameters of our system (the input gains $\alpha$ and $\beta$) must be optimized. In our case, the input mask parameters, namely, the values of $F_1$ in equation (9), or of $F_1$ and $F_2$ in equation (11), as well as the offset $k$ between input and reservoir period in equation (7), must also be tuned in order to find the best system performance. As these parameters influence in a complicated way the dynamics of the reservoir, a systematic sweep over all parameters is carried out.

*4.3. Results when input mask is a single sine*
When investigating the best choice of parameters to build the input mask, we are faced with a problem similar to the one posed by the choice of parameters that describe the reservoir layer. While there is no doubt that the value of the parameters will affect the performance of the system, it is difficult to know beforehand what the best set of parameters will be, and a systematic investigation is usually required. This is due to the fact that the effects of a change in the input mask propagate then through a nonlinear recurrent layer, so it is generally difficult to foresee the effect of a specific input mask on the dynamics of the reservoir layer. However, as in the case of the reservoir architecture, some educated guesses can be made beforehand, to speed up the search of the optimal working point. (For a related discussion see [12]).

As the goal of the input mask is to deliver a different version of the input to each node in the layer, it seems important that the number of nodes $N$ should not have any divisor in common with the $F_1$ coefficient, which represents how many periods of the sine function are contained in the input mask. This guarantees that no two discrete mask coefficients $m_i$, described by equations (10) or (12), have the same value, which clearly is a desirable property of the input mask. Since, historically, the tasks described in section 4.1 of this paper have been tested with $N = 50$, we chose for our experiments the value $N = 53$, in order to be able to compare our results to the ones reported in literature, but also to be able to use all possible values of $F_1$, since $N$ is prime.

In figure 5 we report the result of a study of the influence of the sine period $F_1$ and the loop offset $k$, the parameter used in equation (8) that describes the internal connection of the reservoir layer, in the case where the input mask is a single sine described by equation (10). We measured how the performance of our system on a Channel Equalization task with 32 dB SNR changed as we changed these two parameters.

Two main conclusions can be drawn from this measurement. The first is that there is a complex relation between the value of $F_1$, the value of $k$ and the performance of the system: for a given value of $F_1$ some values of $k$ are "allowed" while some provide worse performance; the same values of $k$ can provide good performance if $F_1$ is changed, and one can distinguish, in dark, bands of "bad" choices of $F_1$ and $k$ in the graph of figure 5. The second conclusion is that the best performance obtained by the single sine mask is still not very good: it is in fact almost two orders of magnitude worse than the one obtained by a randomized mask (see also figure 6).

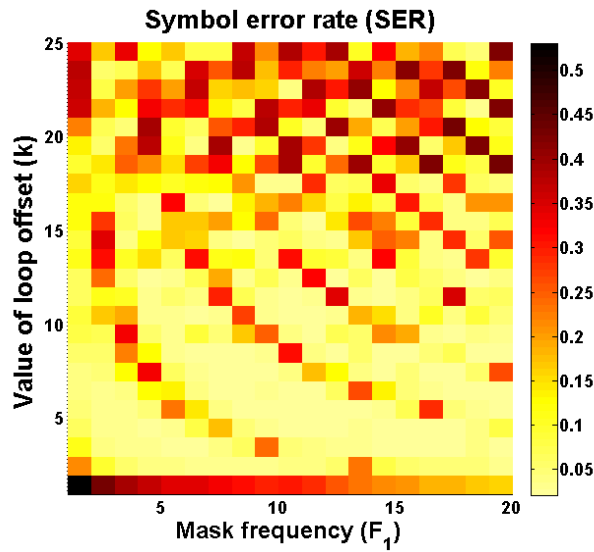

**Figure 5.** Performance (in continuous simulation) of the opto-electronic reservoir when the input mask is a single sine, after optimization over the parameters α, β as a function of the offset k and the frequency $F_1$/T' of the sine. The task used here is the Channel equalization task with 32 dB SNR. The performance is measured by the Symbol Error Rate (SER). Note the complex dependence on the discrete parameters $F_1$ and k.

As the more simplistic approach of a single sine mask failed, we increased the complexity of the input mask by adding a second oscillation, and moved to the two-sine mask described by equation (12). The performance of the reservoir in simulation was considerably better; the simulations showed that the choice of parameters $N = 53$, $F_1 = 3$, $F_2 = 5$, $k = 18$ can provide performances similar to those obtained with a random input mask, as shown in the next section. We note that the best parameter for the input mask obey the requirement of having no common divisors. The choice of a relatively big loop offset $k$ also seems reasonable: a small value of $k$ effectively means that nodes that are connected in the reservoir layer have similar input mask coefficients, while a big value for $k$ helps ensuring that the nodes that share a connection receive a very different version of the input signal. Moreover, as equation (8) shows, a big value of $k$ implies that while the majority of the states at time $n$ are determined by the states of the reservoir nodes at time $n-1$ and the input at time $n$, a significant fraction (actually $k/N$) of the node states depends on the same input at time $n$ but on the state of the reservoir at time $n-2$, thereby probably increasing the complexity of the dynamics in the reservoir layer.

*4.4. Results when input mask is the sum of two sines*

*4.4.1 Channel equalization task*
For the Channel Equalization task on the opto-electronic setup, when the input mask is the sum of two sines as described by equation (11), the results are are, after optimization over all parameters $\alpha$, $\beta$, $k$, $F_1$, $F_2$, comparable to those for a random mask, both in simulation and in experiment, see Figs. 6 and 7. For instance, at SNR=28 dB we obtain a SER of $1.6 \times 10^{-4}$, while the error rate reported in [5] for an optoelectronic reservoir with a random input mask was $1.3 \times 10^{-4}$, and the digital reservoir described in [2] obtains SERs between $10^{-4}$ and $10^{-5}$.

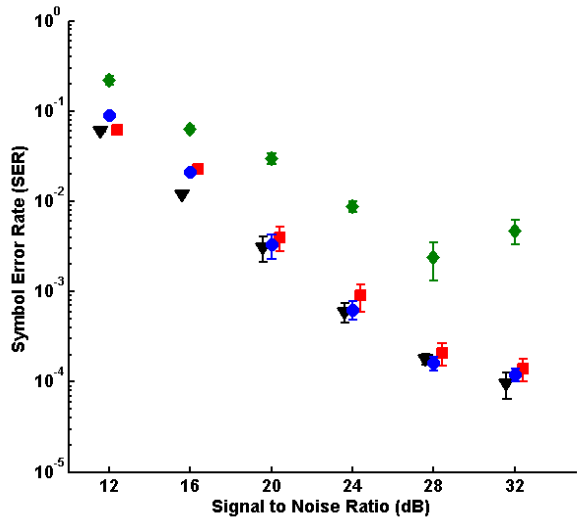

**Figure 6.** Results for the opto-electronic setup on the Channel equalization task. Green diamonds represent the discrete simulation results for input mask consisting of a single sine. Black triangles, red squares and blue circles respectively correspond to discrete simulations, continuous simulations, and experimental results when the input mask consists of the sum of two sines.

We also report in figure 7 a comparison of the experimental results obtained using the opto-electronic and all-optical implementation when the input mask consists of the sum of two sines. This figure shows that the all-optical reservoir performs as well as the optoelectronic one for a signal-to-noise ratio (SNR) ranging from 12 to 16 dB, however for larger SNR its performance is slightly degraded with respect to the all-optical implementation based on a random mask, and even further degraded with respect to the opto-electronic setup. The performance of the optoelectronic setup with a continuous mask is instead very similar to the one obtained with a random mask. Note that in [7] this slight degradation was attributed to spontaneous emission noise. While this remains a possible source of errors, further investigation has also shown the possibility of polarization fluctuations in the fiber spool at a timescale similar to the experiment time. These fluctuations can change the gain of the SOA by about 1 dB. They can undoubtedly affect the performance of the reservoir, since when training the readout weights one of the assumptions is that the dynamical regime of the reservoir layer stays constant for all the duration of the experiment.

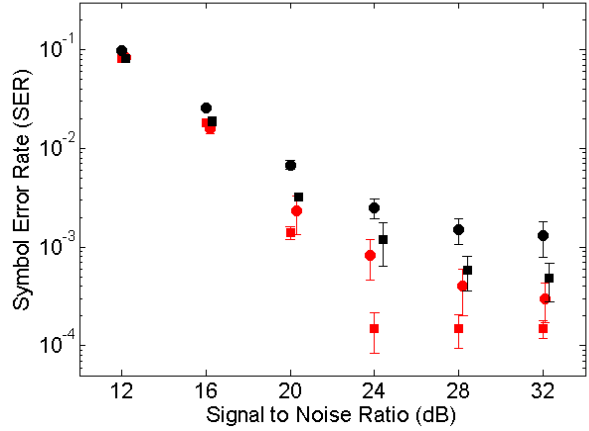

**Figure 7.** Experimental results for opto-electronic and all-optical setups. SER for channel equalization as a function of SNR for opto-electronic (red) and all-optical (black) reservoirs. Circles: continuous input mask consisting of the sum of two sines (the red circle results are the same as those reported in figure 6). Squares: random input mask implemented using digital pre-processing (results taken from our earlier work [5,7]).

*4.4.2 NARMA and memory capacities*
These two tasks are often run in parallel, as they both need a random input sequence $u(n)$ uniformly distributed over some interval; one can then inject an input sequence and then apply several readout weights at the same time, each calculating a different target function, and rescaling the required outputs if needed. The results for these tasks are reported in Table 1

|  | Optoelectronic setup, two-sine mask | Optoelectronic setup, random mask [5] | All-optical setup, two-sine mask | All-optical setup, random mask [7] |
|---|---|---|---|---|
| NARMA10 performance (NMSE) | $0.19 \pm 0.1$ | $0.17 \pm 0.02$ | $0.30 \pm 0.02$ | 0.29 |
| Max. linear memory capacity | $22.1 \pm 0.3$ | 31.9 | $21.8 \pm 0.4$ | 20.8 |
| Max. quadratic memory capacity | $3.2 \pm 0.2$ | 4 | $3.5 \pm 0.05$ | 4.16 |
| Max cross memory capacity | $23 \pm 1.5$ | 27.3 | $13.3 \pm 0.6$ | 8.13 |

**Table 1.** Results for the experimental performance on the NARMA task and the memory capacities for the optoelectronic and the all-optical setups, with a two-sine input mask and a randomized, digitally implemented input mask. Values for the performances of the setups using random masks are taken from our previous works [5,7].

Using a continuous two-sine mask has no negative effect on the performance of the all-optical setup, and some memory capacities appear to be even increased. On the other hand, the linear and cross memory capacities have been slightly decreased for the optoelectronic setup.

*4.4.3 Radar task*
Figure 8 compares the performances of the experimental reservoir computers using a two-sine mask with a digital simulation of a reservoir computer with the same number of nodes $N$ and the same loop offset *k* but a random input mask. The performances reported are relative to two different data series, one generated by high sea waves (average height of 1.8 m) and one by lower sea waves (average height 0.8 m). Obviously the signal generated by rougher sea surfaces is more difficult to predict, especially for long prediction horizons.

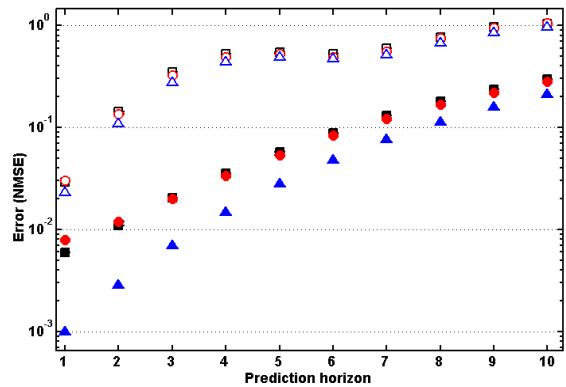

**Figure 8.** Experimental results for opto-electronic and all-optical setups on the radar task. The figure reports the prediction error, measured in NMSE, as a function of the prediction horizon. Full symbols refer to low sea state time series, empty symbols to high sea states. Red circles: optoelectronic setup with two-sine mask, experimental results. Black squares: all-optical setup with two-sine mask, experimental results. Blue triangles: digital, discrete simulation using random input mask and sine nonlinearity.

For this task, the performance on the high sea state time series (upper row, empty symbols) does not depend on the input mask shape, or even on the kind of nonlinearity employed experimentally. The performance appears to be worsened with respect to the case of a random input mask, instead, in the case of low sea state.

*4.4.4 Mackey-Glass time series*
Unlike the other tasks previously described, the Mackey Glass task is one where the performance of the setup with a two-sine mask is greatly inferior to the performance obtained with a random mask, as shown in figure 9.

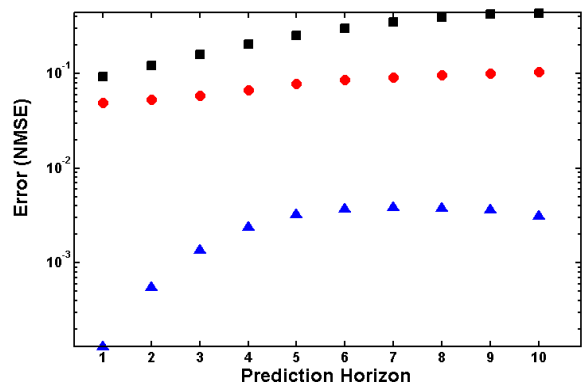

**Figure 9.** Experimental results for opto-electronic and all-optical setups on the Mackey-Glass task. The figure reports the prediction error, measured in NMSE, as a function of the prediction horizon. Red circles: optoelectronic setup with two-sine mask, experimental results. Black squares: all-optical setup with two-sine mask, experimental results. Blue triangles: digital, continuous simulation using random sine mask.

The NMSE is two orders of magnitude higher for the Optoelectronic setup using a two-sine mask than for the simulations run with a random input mask, and the all-optical setup performs even worse.

## 5. Discussion

We have reported the experimental realization of an analog input layer for optical reservoir computers based on the single nonlinear node and delay line architecture. The analog input layer consists in a modulation of the optical signal with a periodic signal (called input mask) before injection into the reservoir. This demonstrates that such reservoir computers can process external signals without the need for digital pre-processing.

We could of course have implemented in these experiments the random input masks used in previous works. However we preferred to study the potential of low complexity masks consisting of one, or the sum of two, sine waves, as such harmonic functions are easy to generated in hardware. As the case of single sine wave did not seem satisfactory, we focused on the case of the sum of two sine waves.

The results on the linear and nonlinear memory tasks show that this implementation can perform complex computations: it can recall inputs from far in the past (large linear memory capacity), and compute nonlinear functions such as the square of an input (quadratic memory capacity), or the product of two inputs (cross memory capacity). Testing on several benchmark tasks shows performance that in some cases is comparable to state of the art, and in other cases is significantly worse. Some insight into when the performance is good, and when it is bad, can be obtained by noting that our system seems to work best for those tasks that do not require very high precision (e.g. channel equalization where the aim is to identify the symbol sent, not reconstruct it exactly; or radar task for the high sea state), and is less satisfactory on tasks that require high precision (radar task for low sea state, Mackey-Glass time series). It would be important to gain further insights into which input masks are appropriate for which tasks. In this respect some of the ideas presented in [12] could be interesting starting points for further investigations.

We expect that further study of analog input layers for reservoir computers will be largely constrained by experimental constraints. Indeed, in a specific physical implementation there will be some parameters of the input mask function that are easy to modify experimentally, and others that are hard to modify (e.g. the amplitude, phase, frequency of sine waves). As the present work shows, the performance of the analog input layer should be studied with respect to these specific parameters, and to the specific tasks of interest. It may be that one can find tradeoffs between simplicity of the input mask and number of nodes in the reservoir.

As a side issue, it should be noted that, unlike the input and reservoir layers, the readout layer presented here is still a digital one, which, in a MATLAB environment, discretizes the value of the internal variables of the reservoir and produces an output in discrete time, as depicted in Figure 1. This approach has been used in all the hardware implementations of reservoir computing, and does not subtract anything from the analog nature of the reservoir layer, and, in our case, of the input mask. Nevertheless, it would of course be extremely interesting to remove this last discretization process, and to be able to build a readout layer that can work in continuous time. The analog readout presented in [13] still relies on the discretization of the reservoir states, so an entirely new architecture is needed. We expect to produce shortly a study in that regard.

In summary, our work demonstrates that in experimental reservoir computers we can replace digital pre-processing by analog input layer, and that a random input mask is not necessary to achieve successful reservoir computing insomuch as high performance on some tasks can be achieved by using a simple input mask that consists of the sum of two sine functions with suitable periods. We believe that the results presented, combined with an analog readout layer similar to the one described in [13], should soon lead to the first fully autonomous high-bandwidth experimental reservoir computer. This would constitute an important milestone in the development of bio-inspired analog computing systems.

ACKNOWLEDGEMENTS
The authors acknowledge financial support from the Fonds de la Recherche Scientifique FRS-FNRS and the Interuniversity Attraction Pole program of the Belgian Science Policy Office under grant IAP P7-35 photonics@be.